\documentclass[preprint,onecolumn,amsmath,amssymb,superscriptaddress,nofootinbib,11pt,a4paper]{revtex4}
\usepackage{url}

\usepackage{epsfig}
\usepackage{amsfonts}
\usepackage{graphicx}
\usepackage{epsfig}
\usepackage{eepic}
\usepackage{amsmath}
\usepackage{amssymb}
\usepackage{color}
\usepackage{bbm}
\usepackage{dcolumn}
\usepackage{bm}
\usepackage{ulem}
\usepackage{mathrsfs}
\usepackage{hyperref}
\usepackage{bbold}
\usepackage{datetime}

\usepackage{overpic} 
\usepackage{rotating}

\usepackage{tikz}
\usepackage{xparse}

\def\bc{\begin{center}}

\def\ec{\end{center}}
\def\be{\begin{eqnarray}}
\def\ee{\end{eqnarray}}
\definecolor{dyellow}{rgb}{1.,0.8,.0}
\definecolor{myblue}{rgb}{.1,.1,.7}
\definecolor{dcyan}{rgb}{.0,.6,.6}
\definecolor{dmagenta}{rgb}{0.6,0.0,0.6}
\definecolor{brown}{rgb}{0.6,0.2,0.}
\definecolor{darkblue}{rgb}{.0,.0,0.5}
\definecolor{darkred}{rgb}{0.75,0.0,0.0}
\definecolor{orange}{rgb}{1.,.6,.0}
\definecolor{dorange}{rgb}{0.8,.4,.0}
\definecolor{darkgreen}{rgb}{0.0,0.6,0.0}
\definecolor{purple}{rgb}{.4,.0,.4}
\definecolor{lightgrey}{rgb}{0.7, 0.7, 0.7}
\definecolor{grey}{rgb}{0.4, 0.4, 0.4}

\usepackage[position=t, singlelinecheck=off]{subfig}
\usepackage[font=small,labelfont=bf,justification=raggedright]{caption}

\def\mW{\mathcal{W}}

\newcommand{\xdownarrow}[1]{%
  {\left\downarrow\vbox to #1{}\right.\kern-\nulldelimiterspace}
}
\newcommand{\xuparrow}[1]{%
  {\left\uparrow\vbox to #1{}\right.\kern-\nulldelimiterspace}
}

\NewDocumentCommand\DownArrow{O{2.0ex} O{black}}{%
   \mathrel{\tikz[baseline] \draw [<-, line width=0.5pt, #2] (0,0) -- ++(0,#1);}
}

\begin{document}
\newsavebox{\lefttempbox}
\title{\bf \Large Topological defects as relics of spontaneous symmetry breaking from black hole physics }

\author{Hua-Bi Zeng}
\affiliation{Center for Gravitation and Cosmology \& College of Physics Science and Technology, Yangzhou University, Yangzhou 225009, China}

\author{Chuan-Yin Xia}
\affiliation{School of Science, Kunming University of Science and Technology,  Kunming 650500, China}
\affiliation{Center for Gravitation and Cosmology \& College of Physics Science and Technology, Yangzhou University, Yangzhou 225009, China}

\author{Hai-Qing Zhang}\email{hqzhang@buaa.edu.cn} {\let\thefootnote\relax\footnotetext{Corresponding author: Hai-Qing Zhang }}
\affiliation{Center for Gravitational Physics, Department of Space Science  \& International Research Institute for Multidisciplinary Science, Beihang University, Beijing 100191, China}

\begin{abstract}
{ 
Formation and evolution of topological defects in course of non-equilibrium symmetry breaking phase transitions is of wide interest in many areas of physics, from cosmology through condensed matter to low temperature physics. Its study in strongly coupled systems, in absence of quasiparticles, is especially challenging. We investigate breaking of U(1) symmetry and the resulting spontaneous formation of vortices in a $(2+1)$-dimensional holographic superconductor employing gauge/gravity duality, a `first-principles' approach to study strongly coupled systems. Magnetic fluxons with quantized fluxes are seen emerging in the post-transition superconducting phase. As expected in type II superconductors, they are trapped in the cores of the order parameter vortices. The dependence of the density of these topological defects on the quench time, the dispersion of the typical winding numbers, and the vortex-vortex correlations are consistent with predictions of the Kibble-Zurek mechanism. }
\vspace{0.5cm}


\end{abstract}

\maketitle



\newpage

{\centering\section{Introduction}}
\label{sec:intro}
Critical dynamics in strongly coupled non-equilibrium phase transitions is one of the most interesting and important problems in modern physics \cite{sachdev}. The conventional quasiparticle approaches do not apply in this case. 
We study the critical dynamics of the superconducting phase transition, focusing on the formation and evolution of topological defects, by utilizing the AdS/CFT correspondence \cite{Maldacena:1997re}. Generation of topological defects is expected in such transitions and can be used to test the Kibble-Zurek mechanism (KZM) \cite{Kibble:1976sj,Kibble:1980mv,Zurek:1985qw,Zurek:1996sj}.

The basic idea of KZM is that, as a system approaches the critical point starting in the symmetric phase, its dynamics undergoes critical slowing down reflected in the divergence of the relaxation time. As a consequence, different domains of the system cannot communicate with each other, and select how to break the symmetry independently. The dimension of these domains is limited by the size of the sonic horizon -- by how far the relevant sound can propagate in the near-critical time interval. 
These independent choices of the broken symmetry lead to irreconcilable differences -- formation of topological defects can be expected, and has been observed. Numerical simulations \cite{Laguna:1996pv,Yates:1998kx,Ibaceta:1998yy,donaire2007,Das:2011cx,Gillman:2017ycq,Antunes:1998rz} as well as experiments in liquid crystals \cite{Chuang:1991zz,Bowick:1992rz,Digal:1998ak}, $^3$He superfluids \cite{Baeuerle:1996zz,Ruutu:1995qz},  Josephson junctions \cite{Carmi:2000zz,Monaco:2002zz,Monaco:2003,Monaco:2005fi}, thin-film superconductors \cite{Maniv:2003zz,golubchik2010,golubchik2011} and quantum optics \cite{Guo} have shown results consistent with KZM (for reviews, see \cite{Kibble:2007zz,delCampo:2013nla}).

Gauge/gravity duality has been employed to study strongly coupled systems to bypass the difficulties caused by the absence of quasiparticles  \cite{Cubrovic:2009ye,Adams:2012pj,Witczak-Krempa:2013nua,Bhaseen:2013ypa}, for a review see \cite{Zaanen:2015oix}. Previous studies on KZM in holographic superfluids and 1D superconducting loop can be found in \cite{Chesler:2014gya,Sonner:2014tca}.  In these two holographic studies, scaling laws between number density of defects and the quench time were found to match KZM in slow quench regime. Other papers on holographic KZM are \cite{Das:2014lda,Natsuume:2017jmu}.

\begin{figure}[h]
\raggedright
{
\includegraphics[trim=0.cm 0cm 0.cm 0.cm, clip=true, height=12cm, width=6cm, scale=0.2, angle=0]{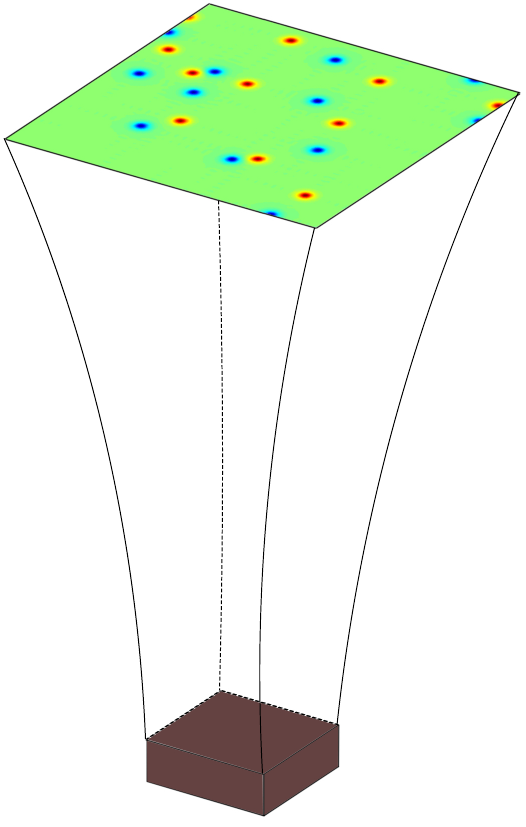}
\put(-190,280){\small\bf(a)}
\put(-120,150){$\rm{\bf AdS}_4: (\Psi, A_\mu)$}
\put(-120,-10){\bf\text{Black}}
\put(-90,-10){\bf\text{Hole}}
\put(-79,333){\begin{rotate}{-19}\bf\text{Boundary}\end{rotate}}
\put(-110,270){\begin{rotate}{45}\tiny\color{purple}\text{quantized vortices}\end{rotate}}
\put(0,0){$\DownArrow[10cm][>=latex]$}
\put(5,15){$z=z_h$}
\put(-3.5,15){$-$}
\put(5,282){$z=0$}
\put(-3.5,282){$-$}
}
\put(50,180){
\includegraphics[trim=.0cm 7cm 0.4cm 7.cm, clip=true, scale=0.4]{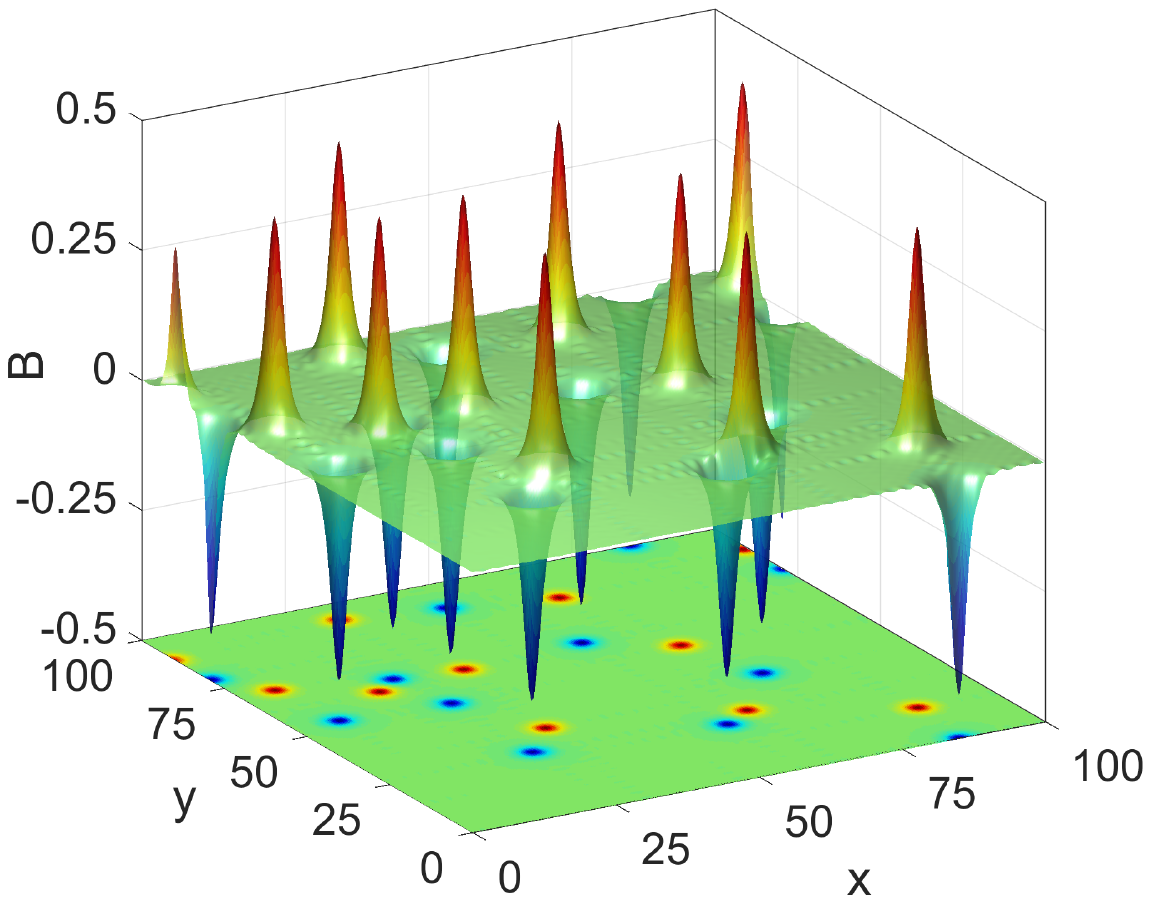}
\put(-235,154){\small\bf(b)}
\put(-125,154){\tiny\color{red}{$\searrow$}}
\put(-143,159){\footnotesize\color{red}{$\Phi\approx1.9955\pi$}}
}
\put(55,-24){
{\includegraphics[trim=4.5cm 9.cm 4.5cm 10cm, clip=true,  scale=0.68]{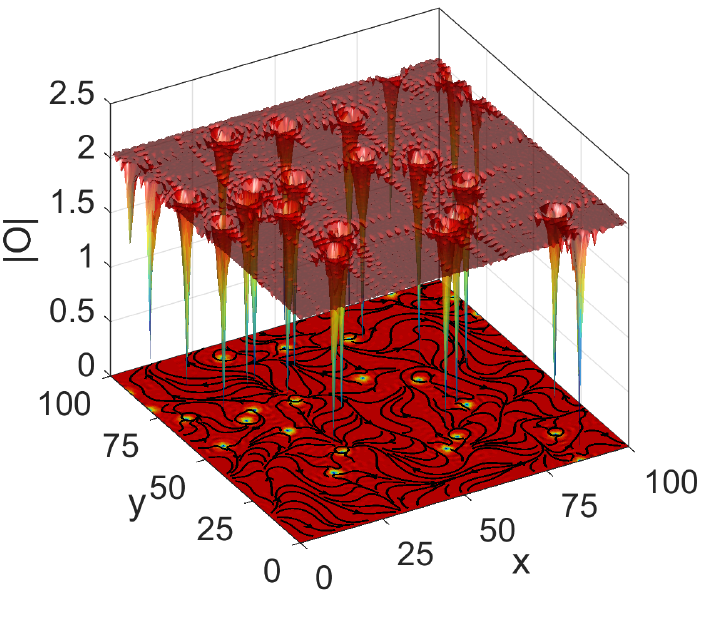}
\put(-235,175){\small\bf(c)}}
}
\caption{\footnotesize{{\bf Holographic setup and birth of topological defects.} {\bf(a)} 
Complex scalar field $\Psi$ and the U(1) gauge field $A_\mu$ are living in the bulk of Schwarzschild-AdS$_4$ spacetime. Quenching temperature across the critical point induces quantized vortices on the (2+1)-dimensional boundary, leading to the holographic KZM.   {\bf(b)} Configurations of the magnetic fluxons and their corresponding locations are shown at the bottom. Red arrow points at a positive magnetic fluxon with magnetic flux $\Phi\approx1.9955\pi$. {\bf(c)} Configurations of the order parameter vortices and the density plot at the bottom.  Streamlines with arrows indicate the directions of phases in the complex plain of the order parameter. Locations of the cores of the vortices and the positions of the magnetic fluxons in panel (b) coincide. For related movie see the \hyperref[app]{Supplemental Information}.
}}
\label{3dvortex}
\end{figure}

We use the AdS/CFT correspondence to examine the U(1) gauge symmetry breaking in the (2+1)-dimensional boundary system, and study the non-equilibrium critical dynamics in a strongly coupled field theory.  The spontaneous generation of magnetic fluxons (with quantized fluxes) is observed for the first time using holographic numerical simulation, Fig. \ref{3dvortex}. Spatial distributions and other characteristics of these topological defects are investigated. This includes their density as a function of the quench time, the dispersion of the typical winding numbers in the resulting superconductor, the flux distribution, and the correlation between the charges and locations of topological defects. We conclude that the resulting defects exhibit the short-range and nearest neighbor vortex-antivortex correlations consistent with the predictions from KZM. 

{ Previous work on holographic KZM did not realize the spontaneously generated magnetic fluxons. For example, in \cite{Chesler:2014gya} the authors studied the holographic KZM for vortices in a superfluid without any magnetic fluxons. In \cite{Sonner:2014tca} the authors studied the holographic KZM in a spatial one dimensional ring which obviously did not have the magnetic fluxons. The authors in \cite{Das:2014lda,Natsuume:2017jmu} analytically investigated the KZM scaling laws nearby the critical point of the phase transition, without any numerical simulations of the vortices, not to mention the magnetic fluxons. The key difference of our paper to previous work is that we realize the spontaneously generated magnetic fluxons in the superconductor system. The advantage of the existence of the magnetic fluxons is that the vortices will exhibit the ``pinning effect'' \cite{tinkham}, which means the vortices will move very slowly in space. Thus, the vortices will be very hard to meet and annihilate. Therefore, the slowing down of the movement will be beneficial to counting the number of vortices, measuring the typical winding numbers and evaluating the vortex-antivortex correlations, which are the tasks in our paper. }
\vspace{1cm}

{\centering\section{Results}}
\label{sect:results}

As observed by Kibble \cite{Kibble:1976sj}, in the early Universe symmetry breaking phase transitions induce local choices of broken symmetry that cannot be coordinated between domains larger than the distance travelled by light since Big Bang till the transition. The resulting a patchwork of domains can then lead to topological defects. 

In the laboratory setting speed of light is no longer relevant. However, critical slowing down that occurs in course of second order phase transitions limits the size of the domains. Equilibrium critical exponents can then predict scaling of the defect density as well as other excitations as a function of the quench time \cite{Zurek:1985qw}.

{\centering\subsection{Magnetic fluxons from symmetry breaking }}
\label{sect:quantum}

Requirements of minimal free energy and periodicity of the phase $\theta$ of the complex scalar field $\Psi$ imply the quantization of the magnetic fluxes generated from U(1) symmetry breaking \cite{tinkham}, i.e., $\Phi=2\pi\mathbb{Z}$, where $\mathbb{Z}$ is an integer. 
Typical configurations of magnetic fluxons and order parameter vortices generated from U(1) symmetry breaking after quench (with quench time $\tau_Q=1000$) are presented in Fig.\ref{3dvortex}(b) and Fig.\ref{3dvortex}(c), respectively.\footnote{We actually quench the charge density on the boundary by fixing the temperature, which is equivalently to quench the temperature. Please see the details in the Methods.} The configurations are in final equilibrium state with temperature $T_f=0.9T_c$. 
The magnetic flux indicated by a red arrow in Fig.\ref{3dvortex}(b) has $\Phi\approx 2 \pi$, which demonstrates the quantization of magnetic flux.  
All ten positive magnetic fluxons shown in this plot have the average $\Phi\approx (1.9864\pm0.0067) \pi $; while the other ten negative magnetic fluxons have average $\Phi\approx (-1.9834\pm 0.0093)\pi$. Therefore, the net magnetic flux of the whole system vanishes within numerical errors, consistent with the fact that there is no external magnetic field imposed for the system. Thus, all magnetic fluxons are {\it spontaneously} generated from  U(1) symmetry breaking due to KZM. \footnote{In fact we made 100 times of simulations for each $\tau_Q$, and the differences of the magnetic fluxes for different runs are very tiny, which are in the range of the order $10^{-2}\pi\sim10^{-3}\pi$. }
At the bottom of Fig.\ref{3dvortex}(c), we show the density plot of the order parameter vortices and the streamlines for the angle of the scalar field phase $\theta$. The arrows indicate the directions of the phase angles. One can read out the positive or negative vorticity of the vortices from the streamlines. Consequently, we see that the locations of positive vortices correspond to positive magnetic fluxons, and vice versa. 




{\centering\subsection{Kibble-Zurek mechanism}}
\label{sec:kzm}

Near the critical point of a second order phase transition, both the relaxation time $\tau$ and the correlation length $\xi$ diverge as,
\begin{eqnarray}\label{cp}
\tau=\tau_0|\epsilon|^{-z\nu},~~~~~~~~
\xi=\xi_0|\epsilon|^{-\nu},
\end{eqnarray}
where $\epsilon=1-T/T_c$ is the reduced dimensionless temperature (or, more generally, dimensionless distance from the critical point), while $\nu$ and $z$ are spatial and dynamical critical exponents. One can usually assume that $\epsilon$ traverses the critical point approximately linearly in time $t$
\be\label{reduceT}
\epsilon(t)=1-{T(t)}/{T_c}={t}/{\tau_Q} \ .
\ee
Above $\tau_Q$ is the quench time. 

KZM recognizes that at the instant $\hat t$ before the critical point, when the rate of change imposed by the quench is comparable to the system's relaxation time $\tau$, the order parameter will cease to follow or even approximate its equilibrium value. Thus, one obtains
\be\label{tfreeze}
{\epsilon(\hat t)}/{\dot\epsilon(\hat t)}=\tau(\hat t)\Rightarrow\hat t=\tau_0^\frac{1}{1+\nu z}\tau_Q^\frac{\nu z}{1+\nu z} \ .
\ee
Time $-\hat t$ marks the beginning of the non-adiabatic evolution and $+\hat t$ its end. What happens inbetween is occasionally described as a ``freeze-out'' but a more accurate picture, especially in the cases when the order parameter is underdamped, is based on the idea of ``sonic horizon'' -- distinct parts of the system can still evolve and influence one another, but they can coordinate their choices of broken symmetry only at distances given by how far the relevant sound (associated with the perturbations of the order parameter) can propagate during the time interval $(-\hat t, +\hat t)$ \cite{Zurek:1985qw}.

Now, from Eq.\eqref{cp} and Eq.\eqref{tfreeze}, one can obtain the corresponding correlation length,
\begin{eqnarray}\label{xifreeze}
\hat\xi=\xi_0\left({\tau_Q}/{\tau_0}\right)^{\frac{\nu}{1+\nu z}}
\end{eqnarray}
The new, post-transition order parameter will randomly select how to break symmetry in domains that are this far apart, as they have no time to communicate with one another. This sonic horizon argument \cite{Zurek:1985qw} leads to domains having the size $\sim \hat\xi$ -- same scaling (although different pre-factors \cite{Francuz:2015zva,Sadhukhan:2019jan}) as these given by the ``freeze-out''.

For vortices in two-dimensional space the estimated number density of point defects is;
\begin{eqnarray}\label{density}
n\propto\hat\xi^{-2}=\xi_0^{-2}\left({\tau_0}/{\tau_Q}\right)^{\frac{2\nu}{1+\nu z}}.
\end{eqnarray}
The same scaling laws (although, again, with somewhat different prefactors) follow from the arguments based on the ``sonic horizon'' paradigm \cite{Zurek:1985qw,delCampo:2013nla,Francuz:2015zva,Sadhukhan:2019jan}. Equations \eqref{tfreeze}, \eqref{xifreeze} and \eqref{density} can be used to test the validity of KZM in laboratory experiments and in numerical simulations.

Another important feature of KZM is the spatial distribution of the charges of topological defects. 
KZM predicts that the random choices of the locally broken symmetry -- e.g., phase of the superfluid wavefunction -- will ultimately lead to anticorrelated charges of topological defects.  This is in contrast to the possibility that topological charges are distributed at random. 

The scalings of the typical winding number $\mW$ subtended by a loop $\cal C$ with circumference $C=2\pi r$ can be used to distinguish between these two alternatives \cite{Zurek:2013qba}. The winding number $\mW$ inside $\cal C$ is $\mW \approx n_+-n_-$ where $n_+$ and $n_-$ are the numbers positively and negatively charged vortices inside $\cal C$. If vortices were distributed at random with randomly assigned topological charges, the dispersion of typical winding numbers $\sqrt{\langle\mW^2\rangle}$ would be proportional to the square root of total number of defects, $n\equiv n_++n_-$ inside the loop. Therefore, it would scale as a square root of the area $A_{\cal C}$ inside $\cal C$, i.e., $\sqrt{\langle\mW^2\rangle}\propto\sqrt{n}\propto\sqrt A_{\cal C}\propto C$. 

However, according to KZM \cite{Zurek:2013qba}, the broken symmetry of the local order parameter -- e.g., phases of the superfluid wavefunction -- rather than topological charge of defects is distributed at random. Therefore, $\mW$ is determined by the winding of the phase along the loop $\cal C$. The random choices of the phase in $\hat \xi$-sized domains along $\cal C$ imply that the accumulated typical winding number should be proportional to  $\sqrt{C/\hat\xi}$ as long as  $C>\hat\xi$, where $C/\hat\xi$ is the number of domains with the size given by the correlation length $\hat\xi$. Therefore, the dispersion of $\mW$ predicted by KZM should scale as,
\be\label{mWC}
\sqrt{\langle\mW^2\rangle}\propto\sqrt{C/\hat\xi}\propto\sqrt{r/\hat\xi}.
\ee
Furthermore, in the range $\langle\mW^2\rangle\gg1$ the absolute value of $\mW$ has the same scaling limit as $\sqrt{\langle\mW^2\rangle}$. More precisely, the relation is \cite{Zurek:2013qba},
\be\label{ww}
\sqrt{\pi/2}~\langle|\mW|\rangle=\sqrt{\langle\mW^2\rangle}
\ee
in which the prefactor $\sqrt{\pi/2}$ comes from the Gaussian approximation to the distribution of $\mW$.

The above scalings need to be adjusted if the magnitude of $\mW$ is smaller than $1$. This happens as the radius $r$ of the contour is smaller than the correlation length $r<\hat\xi$. In this case $\mW$ is proportional to the probability of finding one vortex inside $\cal C$. Thus, $\langle|\mW|\rangle\approx p_++p_-$, where $p_{+/-}$ is the probability to find a positive/negative vortex. In this case $\mW$ only has three possible values $+1, 0$ and $-1$, thus we can deduce $\langle|\mW|\rangle=\langle\mW^2\rangle$. Therefore, in the limit of $\langle|\mW|\rangle\ll1$ we arrive at
 \be\label{rsmall}
 \langle\mW^2\rangle\approx\langle|\mW|\rangle\approx p_{\mW=\pm1}\propto  A_{\cal C}/\hat\xi^2\propto\left(r/\hat\xi\right)^2.
 \ee
 in which $A_{\cal C}$ is the area surrounded by the contour $\cal C$.


\vspace{1cm}
{\centering\subsection{Dynamics of symmetry breaking and nascent topological defects}}
\label{sect:dynamics}

\begin{figure}[h]
\centering
\includegraphics[trim=3.8cm 2.9cm 1.9cm 2.2cm, clip=true, scale=0.67, angle=0]{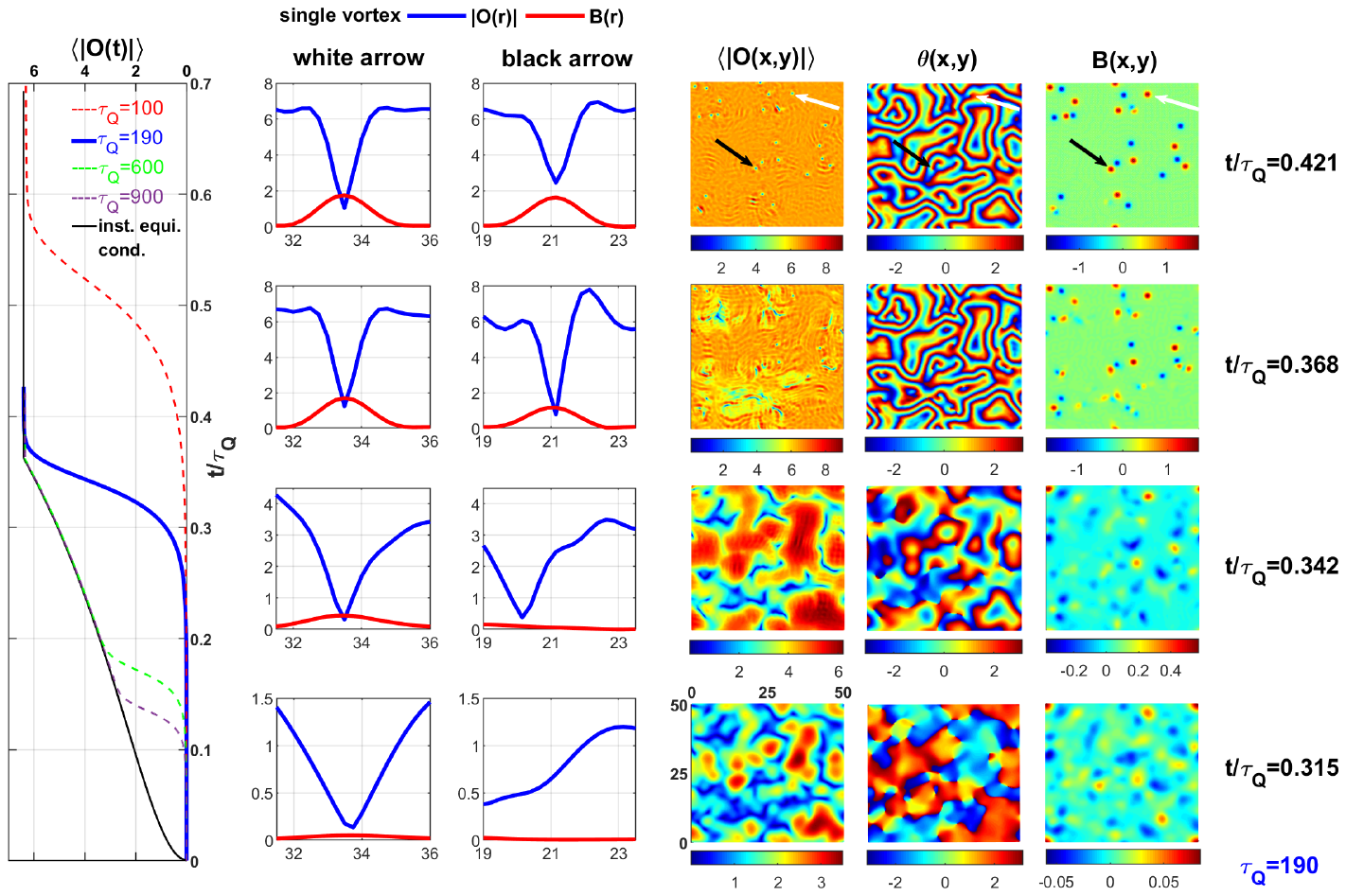}
\put(-430,-10){\small\bf(a)}
\put(-347,-10){\small\bf(b)}
\put(-275,-10){\small\bf(c)}
\put(-200,-10){\small\bf(d)}
\put(-143,-10){\small\bf(e)}
\put(-78,-10){\small\bf(f)}
\caption{\footnotesize{\bf Time evolution of the order parameter and the birth of topological defects.}  {\bf(a)}  The average absolute value of the order parameter $\langle |O(t)|\rangle$ during quenches with $\tau_Q=100$, $190$, $600$ and $900$. Quenches end at final temperature $T_f=0.64T_c$. The black solid line is the instantaneous equilibrium value of the average condensate. Explicit examples of order parameter and magnetic field for $\tau_Q=190$ (thick blue) are shown in the panels {\bf(b)-(f)}.   {\bf(b) $\&$ \bf(c)} Snapshots of the cross sections of condensate and magnetic fields for a single vortex at four specific times with $\tau_Q=190$. Their locations are indicated by the white and black arrows respectively in the subsequent panels {\bf (d)}-{\bf (f)}.  Blue curves are the profiles of the order parameter while red curves are the magnetic fields.  
{\bf(d), (e)} $\&$ {\bf (f)} Density plots of the order parameter $\langle |O(x,y)|\rangle$, phases of the order parameter $\theta(x,y)$ and the corresponding magnetic field $B(x,y)$ at four specific times with $\tau_Q=190$, respectively. 
}\label{vortexTT}
\end{figure}

Growth of the average absolute value of order parameter $\langle | O(t) | \rangle$ from $t=0$ ($T=T_c$) to the final equilibrium state is seen in Fig.\ref{vortexTT}(a).  The instantaneous dynamical values of $\langle | O(t) | \rangle$ remain negligible, i.e. close to what it was in the symmetric vacuum, and, hence, lags behind the instantaneous equilibrium values. For instance for quench with $\tau_Q=190$, its instantaneous value remains negligible until the lag time $t_L/\tau_Q\sim0.263$, and then begins to grow rapidly reaching the approximate equilibrium value at $t/\tau_Q\sim0.368$. This behavior, with $t_L$ larger than but proportional to $\hat t$, is predicted by KZM, and was reported in the past \cite{Das:2011cx,Sonner:2014tca}. 

In Fig.\ref{vortexTT} (b) and Fig.\ref{vortexTT}(c), we track emergence of two vortices for $\tau_Q=190$ at positions $(x,y)\simeq (33,45) $ and $(x,y) \simeq (20,20)$. We only show their cross sections along the $x$-direction in the figure. At time $t/\tau_Q=0.315$ and $t/\tau_Q=0.342$ there are no well-defined vortices as the order parameter is only beginning to grow.  
Well-defined  vortices can be found in the final equilibrium state, for example at $t/\tau_Q=0.421$ in Fig.\ref{vortexTT}(d) and Fig.\ref{vortexTT}(f). 

The top row ($t/\tau_Q=0.421$) in Fig.\ref{vortexTT}(b) and Fig.\ref{vortexTT}(c) demonstrates that the location of the minima of the order parameter coincide with the maxima of the magnetic field. Widths of the flux lines $\lambda_v$ and of the order parameter defects $d_v$ can be fitted by $B(r)\sim e^{-r/\lambda_v}$ and $\langle |O(r)|\rangle\sim\tanh(r/(\sqrt2 d_v))$ \cite{tinkham}. From Fig.\ref{vortexTT}(b) and Fig.\ref{vortexTT}(c), we estimate $\lambda_v\approx 1.15$ and $d_v\approx0.50$, respectively.  Thus, the Ginzburg-Landau parameter is $\kappa=\lambda_v/d_v\approx2.30>1/\sqrt2$, which belongs to type II superconductors \cite{tinkham}.

Fig.\ref{vortexTT}(d) and Fig.\ref{vortexTT}(f) show the density plots of the order parameter $\langle |O(x,y)|\rangle$ and the corresponding magnetic fields $B(x,y)$.  At final equilibrium time $t/\tau_Q=0.421$, positions of the vortices (see the white and black arrows) correspond to the singular points in the density plots of order parameter phases $\theta(x,y)$ in Fig.\ref{vortexTT}(e).  (A movie of the dynamics of the system can be seen in the \hyperref[app]{Supplemental Information}.)

\vspace{1cm}
{\centering\subsection{Number density of topological defects and ``freeze-out'' time}}
\label{app:density}

\begin{figure}[h]
\centering
\captionsetup[subfigure]{labelfont=bf,justification=raggedright}
\subfloat[]{\includegraphics[trim=0cm 0cm 0cm 0cm, clip=true, scale=0.62]{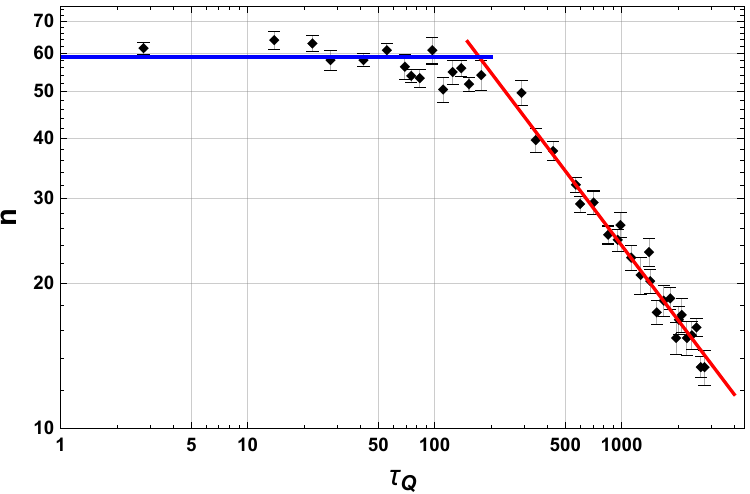}}\quad
\subfloat[]{\includegraphics[trim=0cm 0cm 0cm 0cm, clip=true, scale=0.63]{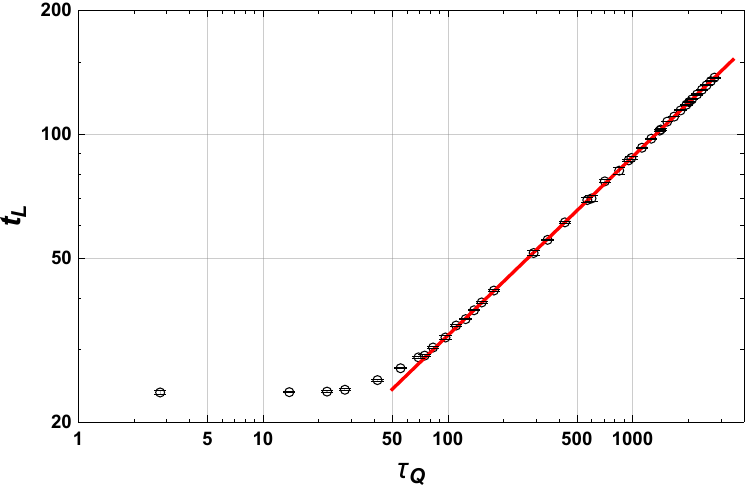}}
\caption{\footnotesize{{\bf Universal scalings of vortex number density and lag time versus quench time.} {\bf(a)} Relation between vortex number density $n$ and quench time $\tau_Q$. Solid diamonds are numerical data while blue and red lines are from the best fit. Error bars denote the standard deviations. Scalings in slow quench (large $\tau_Q$) satisfy KZM prediction, $n\propto\hat\xi^{-2}=\xi_0^{-2}\left({\tau_0}/{\tau_Q}\right)^{\frac{2\nu}{1+\nu z}}$, very well.  
However, vortex number in fast quenches (small $\tau_Q$) is almost constant, independent of $\tau_Q$. These quenches are essentially ``impulse'' -- they start within the impulse interval $(-\hat t, +\hat t)$. {\bf(b)} Relations between the lag time $t_{L}$ and $\tau_Q$. Circles are the numerical data and the error bars are small compared to the size of circles. Red fitted line shows a good agreement with the KZM prediction, 
$t_L \simeq \tau_Q^\frac{\nu z}{1+\nu z} \sim \hat t$} in the slow quench regime. 
}  \label{ntauq}
\end{figure}

We count the vortex number density $n$ as the average order parameter saturates to its equilibrium value.  
Scalings between  $n$ and $\tau_Q$ are exhibited in Fig.\ref{ntauq}(a), with the final equilibrium temperature $T_f=0.64T_c$ and the size of the boundary $(x,y)=(100,100)$. In the slow quench regime (large $\tau_Q$), the scaling relation is fitted as $n\approx (836.0430\pm1.1844)\times\tau_Q^{-0.5126\pm0.0230}$,  where the uncertainties give standard deviations. The quasi-normal modes (QNMs) analysis in the \hyperref[app]{Supplemental Information} indicates a mean-field theory with $\nu\approx1/2$ and $z\approx2$. Thus, the exponent in the scaling between $n\sim\tau_Q$ in Eq.\eqref{density} is roughly $-1/2$, which is in good agreement with the above numerical results. By contrast, in the fast quench regime (small $\tau_Q$ which is beyond the scope of KZM) the vortex number is approximately constant.  This is consistent with the previous results \cite{Chesler:2014gya, Sonner:2014tca, Kibble:2007zz}.

The time $\hat t \sim \sqrt {\tau_Q}$ in the symmetry breaking phase is the instant that the system leaves the phase of the quench when the dynamics of the systems cannot keep up with the changes imposed by the quench and enters the adiabatic region \cite{delCampo:2013nla}. Following \cite{Das:2011cx,Sonner:2014tca} we define the lag time $t_L$ as the time when the order parameter begins to grow rapidly. 
Lag time $t_L$ reflects the ``freeze-out'' time $\hat t$ \cite{Das:2011cx,Sonner:2014tca}.   
Scalings between $t_{L}$ and $\tau_Q$ are shown in Fig.\ref{ntauq}(b). The best fit in the slow quench regime (red line) is $t_L\approx (4.3345\pm0.0205)\times\tau_Q^{0.4860\pm0.0007}$, in which the exponent matches KZ relation when we take $\nu=1/2$ and $z=2$ in Eq.\eqref{tfreeze}.

{\centering\subsection{Typical winding number $\mW$ }}
\label{sect:distribution}

The winding number $\cal W$ is phase accumulated along the contour $\cal C$. In the broken symmetry phase it is dominated by the net number of topological defects inside $\cal C$. According to KZM, random choices of the phase of the order parameter determine the distribution of the topological defects. As a consequence, defect charges are not distributed randomly, but, rather, anticorrelated. This leads to predictions about the distribution of winding numbers as a function of the circumference of $\cal C$ \cite{Zurek:2013qba}.
\begin{figure}[h]
\centering
{\includegraphics[trim=4.3cm 9.4cm 4.2cm 9.8cm, clip=true, scale=0.55]{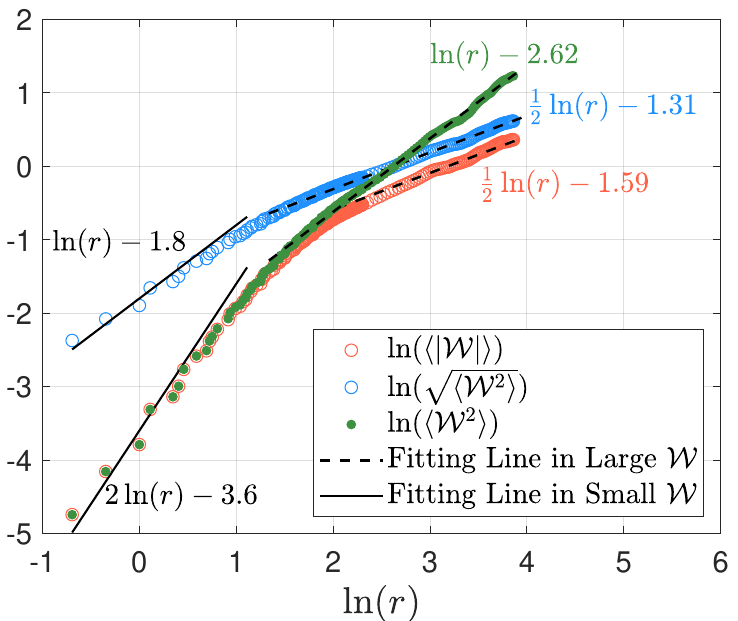}
\put(-210,155){\small\bf(a)}}~~~~~~~
{\includegraphics[trim=4.3cm 9.4cm 3.7cm 9.8cm, clip=true, scale=0.55]{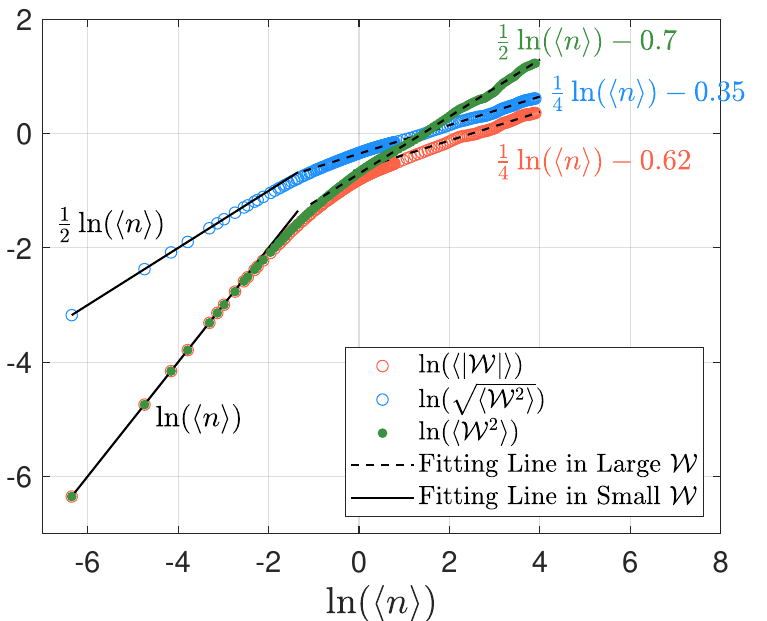}
\put(-218,155){\small\bf(b)}}
\caption{\footnotesize{\bf Universal scalings of typical winding number $\mW$.} {\bf (a)} Typical winding number $\mW$ as a function of the size $r$ of the contour $\cal C$ in which the defects are trapped. {\bf (b)} Scalings between $\mW$ and average total number of defects $\langle n\rangle$ trapped inside contour $\cal C$. In panels (a) and (b) the data points were obtained by averaging 200000 runs for parameters {$\tau_Q=30$} and $T_f=0.64T_c$. Dots and open circles are the numerical data while the solid and dashed black lines are the fittings of the data in the small and large $\mW$ limits, respectively.  }\label{wzurek}
\end{figure}

Fig.\ref{wzurek}(a) shows the relation between three functions of $\mW$ to the radius $r$ of the contour $\cal C$ (with circumference $C\propto r$), inside which vortices are trapped. 
As $r$ is large (or $C>\hat\xi$), the dispersion of $\mW$ is proportional to the square root of $r$ as is shown in Eq.\eqref{mWC}. This scaling is reflected in our holographic results in Fig.\ref{wzurek}(a) with $\ln(\sqrt{\langle\mW^2\rangle})\approx 1/2\ln(r)-1.31$.   
From Gaussian approximation to the distribution of $\mW$, one can deduce the relation $\sqrt{\pi/2}\langle|\mW|\rangle=\sqrt{\langle\mW^2\rangle}$ when $\mW$ is large, as is indicated in Eq.\eqref{ww}. This relation is also verified in Fig.\eqref{wzurek}(a) with the ratio $\sqrt{\langle\mW^2\rangle}/\langle|\mW|\rangle\approx { e^{-1.31}/e^{-1.59}\approx1.3231}$  or using Fig.\eqref{wzurek}(b) where $\sqrt{\langle\mW^2\rangle}/\langle|\mW|\rangle\approx { e^{-0.35}/e^{-0.62}\approx1.3099}$, compared to the theoretical value $\sqrt{\pi/2}\approx 1.2533$.  
In the opposite limit $|\mW|\ll1$, Eq.\eqref{rsmall} is verified as well using holography as we examine the scalings of $\langle\mW^2\rangle$ and $\langle|\mW|\rangle$ in Fig.\ref{wzurek}(a) that $\ln(\langle\mW^2\rangle)\approx\ln(\langle|\mW|\rangle)\approx2\ln(r)-3.6$.

Fig.\ref{wzurek}(b) exhibits the relations between $\mW$ and the average total vortex number $\langle n\rangle$ inside the contour $\cal C$. Because of $\langle n\rangle\sim  A_{\cal C}/\hat\xi^2\propto r^2/\hat\xi^2$, the doublings of powers for scalings $\mW\sim\langle n\rangle$ and $\mW\sim r$ are explicitly shown by comparing Fig.\ref{wzurek}(b) and Fig.\ref{wzurek}(a). In the limit $r<\hat\xi$, it is usually only possible to find one vortex (either positive or negative) inside $\cal C$, thus $\langle|\mW|\rangle\approx\langle\mW^2\rangle\approx \langle n\rangle$ inside a small $\cal C$. This is demonstrated as well -- $\ln(\langle|\mW|\rangle)\approx\ln(\langle\mW^2\rangle)\approx\ln(\langle n\rangle)$ in Fig.\ref{wzurek}(b) when $\langle n\rangle$ is small. 

Therefore, the universal scalings of typical winding number $\mW$ in our holographic study are consistent with predictions from KZM \cite{Digal:1998ak,Zurek:2013qba,Lin:2015lca}.

{\centering\subsection{Vortex-vortex correlation function with polarity}}
\label{sect:vvcorrelation}

Correlation length $\hat\xi$ can be estimated from the above $\mW$. We see that, $\langle|\mW|\rangle=\langle\mW^2\rangle$ as $r<\hat\xi$, however, this equality is violated as $r\gg\hat\xi$. Fig.\ref{wzurek}(a) shows that $\langle|\mW|\rangle$ departs from $\langle\mW^2\rangle$ at around $1<\ln(r)< 2$, thus we can estimate $2.7<\hat\xi< 7.3$. 

\begin{figure}[h]
\centering
{\includegraphics[trim=1.8cm 7.4cm 1.7cm 6.5cm, clip=true, scale=0.5]{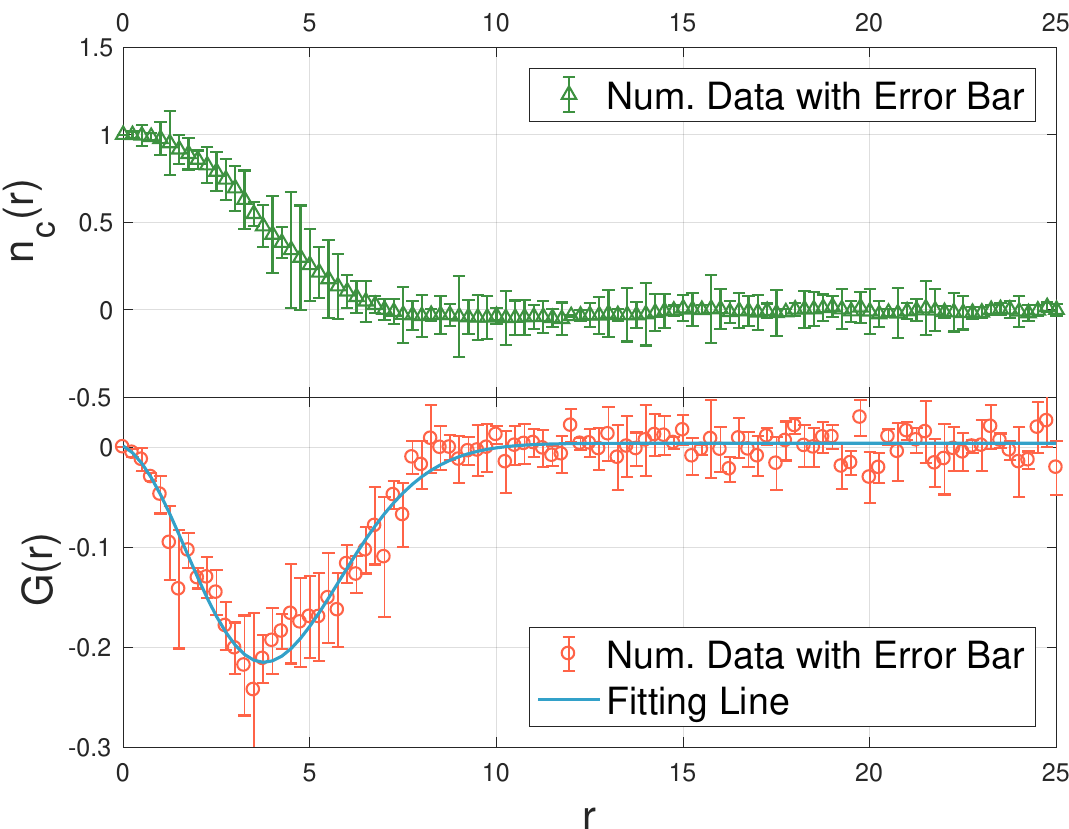}}
\put(-270,185){\small\bf(a)}
\put(-270,92){\small\bf(b)}
\caption{\footnotesize{{\bf Net vortex number $n_c(r)$ and vortex-vortex correlation function $G(r)$.}  {\bf(a)} Profile of net vortex number $n_c(r)$. It decreases from $1$ at $r=0$ to $0$ at large $r$, indicating vortex-antivortex correlations. Error bars stand for the standard deviations. {\bf(b)} Profile of vortex-vortex correlation function $G(r)$. Minimum of the fitting curve indicates the correlation length as $\hat\xi\approx 3.7712$. Parameters $\tau_Q$ and $T_f$ are the same as FIG.\ref{wzurek}. }}\label{vvcorre}
\end{figure}


Following \cite{golubchik2010,golubchik2011}, one can evaluate $\hat\xi$ from the vortex-vortex correlation function $G(r)$ with vortex polarities as well.  $G(r)$ is defined as $G(r)\equiv\langle n(r)n(0)\rangle$, with $n(r)=+1/-1$ at the location of a positive/negative vortex, and $0$ elsewhere. In practice, $G(r)$ can be evaluated by summing over all charges of vortices (with $\pm$ polarities) sitting at the circumference a contour $\cal C$, whose center is at a positive vortex.   
Meanwhile, one can also define the net vortex number $n_c(r)$ by summing over vortex charges inside the above contour\footnote{Please note the different definitions of $n_c$ and the aforementioned typical winding number $\mW$. $\mW$ is defined as the net vortex number inside a contour whose center can be anywhere; $n_c$ is defined inside a contour whose center is only at a positive vortex.}.   
Fig.\ref{vvcorre}(a) shows that $n_c=1$ at $r=0$, which is obvious from its definition. Away from $r=0$, $n_c$  decreases to zero, which demonstrates the short-range vortex-antivortex correlations.    
From the definition of $G(r)$, one can set $G(0)=0$.  Negative minimum of $G(r)$ in Fig.\ref{vvcorre}(b) also reflects the short-range, nearest neighbour vortex-antivortex correlations between vortices, and the position of the minimum is $\hat\xi$.  
Fitting $G(r)\approx a r^2 \times e^{-b r^2}$ to the theory \cite{halperin,liu-mazenko}, we get $ a\approx -0.0412$ and $b\approx 0.0703$.  Therefore, the correlation length can be estimated as $\hat\xi\approx1/\sqrt{b}\approx3.7$, which lies in the range $2.7<\hat\xi< 7.3$ of our previous estimate. 

There is another length scale -- mean vortex separation $r_{av}=\sqrt{A/\langle n\rangle}$, in which $A$ is the area of the system \cite{golubchik2010}. Thus, $r_{av}\approx12.9099$ as $\tau_Q=30$ from Fig.\ref{ntauq}(a)  (with $A=100\times100, \langle n\rangle\approx60$). Consequently, $\hat\xi\approx0.2921 r_{av}$ which is comparable to the experimental results $\hat\xi\approx0.35r_{av}$ in \cite{golubchik2010}, where the authors studied the distributions of the magnetic flux quanta from KZM in a 2D superconducting film.

{\centering\section{Conclusions}}
\label{sect:discussions}
Taking advantage of the AdS/CFT correspondence we have simulated quench-induced symmetry breaking in the transition from the normal to superconducting phase of the strongly coupled holographic field theory. We have observed formation of topological defects -- fluxons with quantized fluxes trapped within the vortices, with properties consistent with type II superconductor. Their densities accord with the predictions of KZM, as does their distribution. In particular, they are anticorrelated. This is related to the distribution of the winding numbers of the phase of the condensate. They provide evidence that it is the condensate phase (i.e., post-transition choice of the broken symmetry) that is random, which in turn results in the anticorrelation of the topological charges. We have also observed that the lag time -- the instant at which the order parameter begins to grow rapidly -- scales as the freezeout time $\hat t$, central to KZM.

\vspace{1cm}
{\centering\section*{Methods}}
\label{sect:methods}
\phantomsection

{\bf Holographic Setup}:  In the probe limit, we adopt the black brane background as
\begin{equation}
ds^2 = \frac{L^2}{z^2} (-f(z) dt^2 - 2dtdz + dx^2 + dy^2),
\end{equation}
where $f = 1 - (z/z_h)^3$. The location of horizon is at $z_h$ while $z = 0$ is the boundary where the field theory lives. The Hawking temperature of the black brane is $T=3/(4\pi z_h)$ which also corresponds to the temperature of the dual field theory. In the numerics we have scaled $z_h=1$.   We take the commonly used Einstein-Maxwell-complex scalar model in a holographic superconductor \cite{hartnoll},
\begin{equation}\label{lag}
\mathcal{L} = -\frac{1}{4} F_{\mu \nu} F^{\mu \nu} - |D \Psi|^2 - m^2 |\Psi|^2.
\end{equation}
where $D=\nabla -iA$. (Throughout this paper, we work in the units with $e=c=\hbar=k_B=1$.) The ansatz we  take is $\Psi = \Psi(t,z,x,y)$, $A_t = A_t(t,z,x,y)$, $A_x = A_x(t,z,x,y)$, $A_y = A_y(t,z,x,y)$ and $A_z = 0$. Then the equations of motion (EoM) read,
\begin{eqnarray}\label{eomofwhole}
D_\mu D^\mu\Psi-m^2\Psi=0,~~~\nabla_\mu F^{\mu\nu}=i\left(\Psi^* D^\nu\Psi-\Psi{(D^\nu\Psi)^*}\right),
\end{eqnarray}

The asymptotic expansions of fields near the boundary $z\to0$ are  (we have set $m^2= -2/L^2$)
\be
A_\mu\sim a_\mu+b_\mu z+\dots,~~~\Psi=\frac zL\left(\Psi_0+\Psi_1 z+\dots\right)
\ee
 In the numerics we have scaled $L=1$. From AdS/CFT correspondence, $a_t, a_i~ (i=x, y)$ and $\Psi_0$  are interpreted as the chemical potential, gauge field velocity and source of scalar operators on the boundary, respectively. Their corresponding conjugate variables can be achieved by varying the renormalized on-shell action $S_{\rm ren.}$ with respect to these source terms.  From holographic renormalization \cite{Skenderis:2002wp}, we can add the counter terms of the scalar fields $S_{\rm c.t.}=\int d^3x\sqrt{-\gamma}\Psi^*\Psi$ into the divergent on-shell action, where $\gamma$ is the determinant of the reduced metric on the $z\to0$ boundary. In order to get the dynamical gauge fields in the boundary, we impose the Neumann boundary conditions for the gauge fields as $z\to0$ \cite{witten,silva}. Thus, the surface term $S_{\rm surf.}=\int d^3x\sqrt{-\gamma}n^\mu F_{\mu\nu}A^\nu$ near the boundary should be added as well in order to have a well-defined variation, where $n^\mu$ is the normal vector perpendicular to the boundary. Hence, we obtain the finite renormalized on-shell action $S_{\rm ren.}$. Therefore, the expectation value of the order parameter,  $\langle O\rangle=\Psi_1$, can be obtained by varying $S_{\rm ren.}$ with respect to $\Psi_0$. Expanding the $z$-component of Maxwell equations near boundary, we get $\partial_tb_t+\partial_iJ^i=0$, which is exactly a conservation equation of the charge density and current on the boundary, since from the variation of $S_{\rm ren.}$ one can easily deduce that $b_t=-\rho$ with $\rho$ the charge density and $J^i=-b_i-(\partial_ia_t-\partial_ta_i)$ which is the $i$-direction current respectively.

On the boundary, we set $\Psi_0=0$ in order to have spontaneously broken symmetry of the order parameter. The Neumann boundary conditions for the gauge fields can be imposed from the above conservation equations. Therefore, dynamical gauge fields on the boundary can be computed and lead to the spontaneous formation of magnetic vortices. Moreover, we impose the periodic boundary conditions for all the fields on the spatial boundary along $(x, y)$-directions. Near the horizon we set $A_t(z_h)=0$ and the regular finite boundary conditions for other fields.

From the dimension analysis, we know that the temperature of the black hole $T$ has mass dimension one, while the mass dimension of the charge density $\rho$ has mass dimension two. Therefore, $T/\sqrt{\rho}$ is dimensionless. From holographic superconductor \cite{hartnoll}, decreasing the temperature is equivalent to increasing the charge density. Therefore, in order to linearize the temperature near the critical point like Eq.\eqref{reduceT}, we can actually quench the charge density $\rho$ as $\rho(t)=\rho_c\left(1-t/\tau_Q\right)^{-2}$ with $\rho_c\approx4.06$, {which is the critical charge density from the equilibrium state of the holographic superconductor. For all quench rates in our paper, we quench the temperature from the initial temperature $T_i=1.30T_c$ to the final temperature $T_f=0.64T_c$ (Except in Fig.\ref{3dvortex}, the final temperature is $T_f=0.9T_c$). After the quench, we maintain the temperature at $T_f$ until the system arrives at the final equilibrium state. }

{\bf Numerical Schemes:}  
We thermalize the system thoroughly before quench in order to make an equilibrium initial state.
In order to thermalize the system initially, different from putting the seeds on the boundary in \cite{Chesler:2014gya,Sonner:2014tca}, we put the random seeds of the fields in the bulk by satisfying the statistical distributions $\langle s(t,x_i)\rangle=0$ and $\langle s(t,x_i)s(t',x_j)\rangle=h\delta(t-t')\delta(x_i-x_j)$, with the amplitude  $h=10^{-3}$. \footnote{Other relatively smaller magnitudes of $h$ lead to similar results. In principle, $h$ cannot be too large since the seeds serve as perturbations to thermalize the system.}  System evolves by using the fourth order Runge-Kutta method with time step $\Delta t=0.025$. In the radial direction $z$, we use the Chebyshev  pseudo-spectral method with 21 grids. Since in the $(x, y)$-directions, all the fields are periodic, we use the Fourier decomposition along $(x, y)$-directions {with the spatial spacing $\Delta x=\Delta y=0.25$. The codes are indeed very robust to the grids in time and spatial directions. We choose these grids in our paper considering both the consumptions of time and the good quality of the results.} Filtering of the high momentum modes are implemented following the ``$2/3$'s rule'' that the uppermost one third Fourier modes are removed \cite{Chesler:2013lia}.

\vspace{1cm}
{\centering\section*{Acknowledgment}}
 The authors greatly thank the comments, discussions and collaborations from Wojciech H. Zurek during the completion of this work. The authors will also thank Adolfo del Campo, Victor Cardoso, Chiang-Mei Chen, Antonio M. Garc\'ia-Garc\'ia, Daniel Golubchik, Sean Hartnoll, Zhi-Hong Li, Peng Liu, Juan Maldacena, Julian Sonner, Yu Tian, Edward Witten and Jan Zaanen for helpful discussions. H.B.Z. and H.Q.Z. are supported by the National Natural Science Foundation of China (Grants No. 11675140, 11705005 and 11875095). 

\vspace{1cm}
{\centering\section*{Author contributions}} 
H.B.Z. and  H.Q.Z. conceived the research. H.B.Z., C.Y.X. and H.Q.Z. contributed equally to the writings of the codes. H.B.Z. and H.Q.Z. analyzed the data and contributed to the writings of the paper. 

\vspace{1cm}
{\centering\section*{Competing interests}}
The authors declare no competing interests.

\newpage
\pagebreak
\clearpage
\label{app}

\begin{center}
\textbf{\large ---Supplemental Information---}
\end{center}


\setcounter{equation}{0}
\setcounter{figure}{0}
\setcounter{table}{0}
\setcounter{section}{0}
\setcounter{page}{1}
\makeatletter
\renewcommand{\theequation}{S\arabic{equation}}
\renewcommand{\thefigure}{S\arabic{figure}}
\renewcommand{\bibnumfmt}[1]{[S#1]}
\renewcommand{\citenumfont}[1]{S#1}

{\large\section{Movies of the dynamics of the system}}
\label{app:dyn}

\begin{figure}[h]
\centering
\includegraphics[trim=3.2cm 9.7cm 3.cm 9.5cm, clip=true, scale=1.1, angle=0]{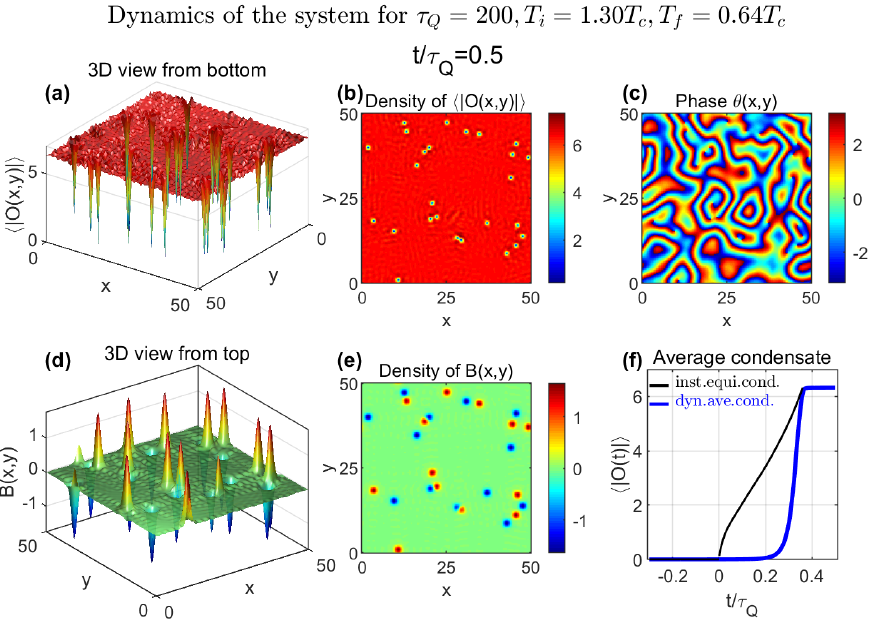}
\caption{\footnotesize{{\bf Snapshot from the Supplementary Movie at the final equilibrium time.} {\bf (a)} Three dimensional visualization of the order parameter condensate on the boundary. {\bf (b)} Density plot of the order parameter condensate. {\bf(c)} Density plot of the phase of the order parameter. {\bf(d)} Three dimensional visualization of the magnetic field on the boundary. {\bf (e)} Density plot of the magnetic field. {\bf (f)} Time evolution of the instantaneous equilibrium condensate (black line) and the dynamical average condensate (blue line).   }}\label{supmovie}
\end{figure}

A movie of the time evolution of the system can be found in the Supplementary Movie (or refer to it from this link \href{https://bhpan.buaa.edu.cn:443/link/CC186969CE471365E3A3CABE74B04CB7}{Supplementary Movie}), in which the quench time is $\tau_Q=200$, the initial temperature is $T_i=1.30T_c$ and the final equilibrium temperature is $T_f=0.64T_c$. In this movie, the animation $\bf(a)$ is three dimensional visualization of the dynamics of the order parameter condensate $\langle|O(x,y)|\rangle$ on the boundary; The animation $\bf(b)$ is about the density of the condensate $\langle|O(x,y)|\rangle$  while the animation $\bf(c)$ shows the density of the phase $\theta(x,y)$ of the order parameter; The animation $\bf(d)$ exhibits the dynamics of the magnetic field $B(x,y)$ on the boundary while the animation $\bf(e)$ is its corresponding density; The last animation $\bf(f)$ contains the time evolution of two kinds of average condensates: one is the black line which is the instantaneous equilibrium condensate (inst.equi.cond.),  the other one is the blue line which indicates the dynamical average condensate (dyn.ave.cond.). As an example, in Fig.\ref{supmovie} of this Supplementary Information we show the snapshot from the movie at the final equilibrium time.

\vspace{1cm}
{\large\section{Quasi-normal modes}}
\label{app:qnm}

\begin{figure}[h]
\centering
\captionsetup[subfigure]{position=top,labelfont=bf,singlelinecheck=off,justification=raggedright}
\subfloat[]{\includegraphics[trim=1.8cm 6.5cm 2.4cm 7.5cm, clip=true, scale=0.45, angle=0]{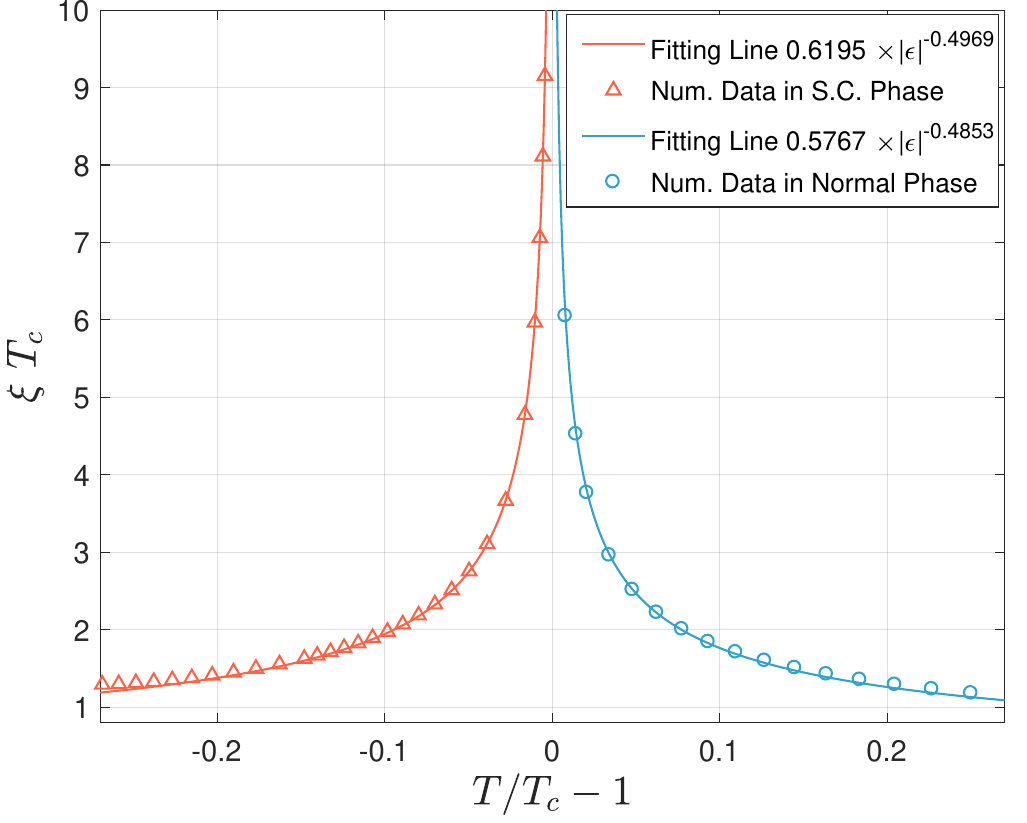}}~~
\subfloat[]{\includegraphics[trim=1.6cm 6.5cm 2.4cm 7.5cm, clip=true, scale=0.45, angle=0]{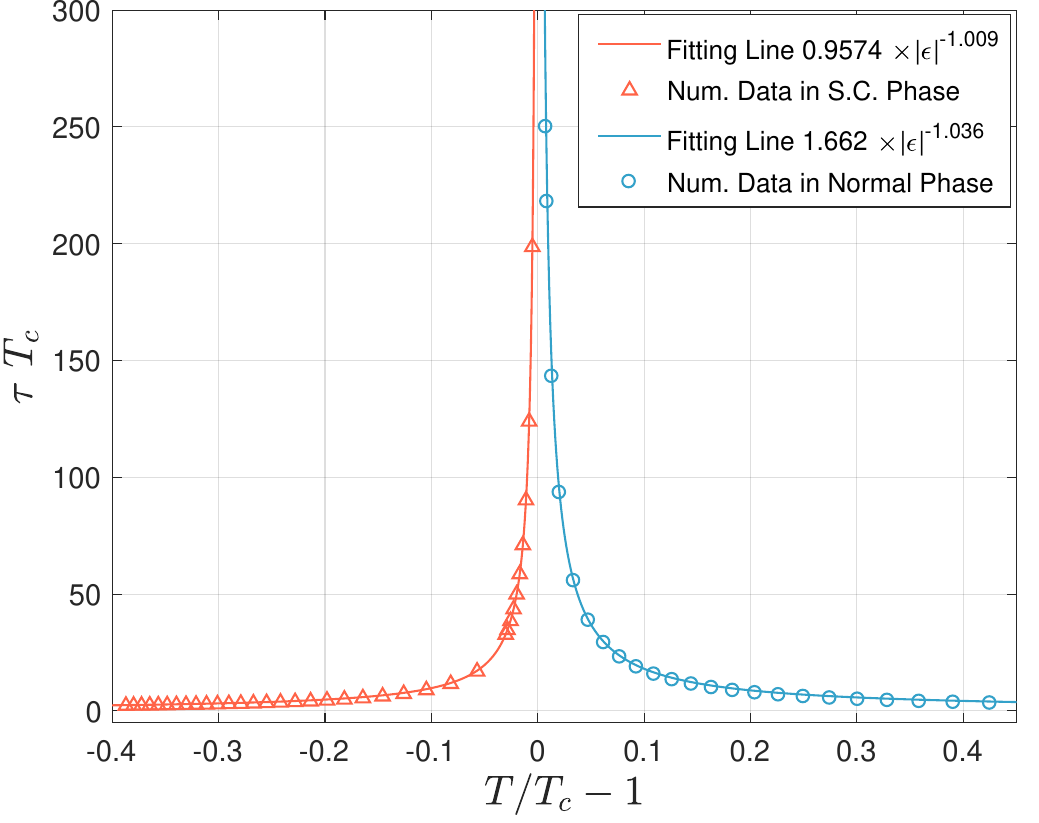}}
\caption{\footnotesize{{\bf Temperature dependence of correlation length and relaxation time in the normal and superconducting state from QNMs.} {\bf(a)} The dimensionless correlation length $\xi T_c$ vs. the reduced temperature $\epsilon=1-T/T_c$. The red line and triangles are for the superconducting phase, while the blue ones are for the normal state. The critical exponents are close to $-1/2$, which indicates a mean field theory; {\bf(b)} The dimensionless relaxation time $\tau T_c$ as functions of $\epsilon$. The red line and triangles are for the superconducting phase, while the blue ones are for the normal state. The critical exponents are close to $-1$, which is consistent with the mean field theory.}}\label{thermal_xi}
\end{figure}

From holography, QNMs correspond to the poles of the correlation functions \cite{Kovtun:2005ev}. We can read off the correlation length $\xi$ from the correlation function as \cite{Maeda:2009wv}
\begin{equation}\label{chiwk}
\langle O(\omega, k)O^\dagger(-\omega, -k)\rangle \sim \frac{1}{i\tilde c \omega+k^2+1/\xi^2}.
\end{equation}
where $k$ and $\omega$ are the momentum and frequency of the modes of perturbations respectively, and $\tilde c$ is a parameter. By linearly perturb the equations of motions in Schwarzschild-AdS metric, we are able to calculate the QNMs of the scalar fields.

In the normal state, $A_t$ is the background gauge field, while $\psi$ is the first order perturbations of the scalar field $\Phi$ as $\delta\Phi=\psi(z)\exp{(-i\omega t+ikx)}$ and $\psi$ is real {(Note that we have scaled $\Phi=\Psi/z$). } The only one decoupled EoM is 
\be\label{normalscalar}
\psi''+\frac{f'}{f}\psi'+\left(\frac{(A_t+\omega )^2}{f^2}+\frac{z f'-k^2z^2+2}{z^2 f}-\frac{2}{z^2}\right)\psi
=0
\ee

In the superconducting state with $\Psi\neq0$, one needs to construct gauge-invariant perturbations since there are mixed perturbations in this case. The infinitesimal gauge transformations are \cite{Amado:2009ts}
\be
\{\delta a_t, \delta a_x, \delta a_z\}&=&\{-i\omega\lambda, ik\lambda,\lambda'(z)\}, \\
\{\delta\sigma, \delta\eta\}&=&\{0, \lambda \Phi\}.
\ee
in which $\sigma$ and $\eta$ are respectively the real and imaginary parts of perturbations of the scalar field, while $a_i (i=t, x, z)$ are the perturbations of the corresponding gauge fields.  
Thus, the independent gauge-invariant quantities can be built as
\be
\Phi_1=\sigma,\ \Phi_2=i\omega\eta+\Phi a_t,\ \Phi_3=-ik\eta+\Phi a_x.
\ee

 For $k=0$, the equations for the gauge-invariant quantities are
\be\label{k0scalar1}
{\Phi_1} \left(\frac{{A_t}^2+\omega^2}{f^2}+\frac{f'}{z
   f}+\frac{2}{z^2 f}-\frac{2}{z^2}\right)+\frac{2 {A_t} }{f^2}{\Phi_2}+\frac{f' }{f}{\Phi_1}'+{\Phi_1}''&=&0, \\
{\Phi_2} \left(\frac{{A_t}^2+\omega ^2}{f^2}+\frac{z
   f'-2 z^2 {\psi_0}^2+2}{z^2 f}-\frac{2 {\psi_0}'
   \left({\psi_0} f'+2 f{\psi_0}'\right)}{\omega
   ^2-2 f {\psi_0}^2}-\frac{2}{z^2}\right)&&\nonumber\\
   +\frac{2 {A_t}
   {\Phi_1} \left(\omega ^2-2 f {\psi_0}^2\right)}{f^2}+\frac{{\Phi_2}' \left(\omega ^2 f'+4
   f^2 {\psi_0} {\psi_0}'\right)}{\omega ^2 f-2
   f^2 {\psi_0}^2}+{\Phi_2}''&=&0\label{k0scalar2}
\ee
in which the equations of $\Phi_3$ is  decoupled.

For $\omega=0$, the equations for gauge-invariant perturbations become
\be\label{w0scalar1}
\Phi_1 \left(\frac{A_t^2}{f^2}+\frac{z f'-k^2 z^2 -2 f+2 }{z^2f}\right)+\frac{2 A_t }{f^2}\Phi_2+\frac{f' }{f}\Phi_1'+\Phi_1''&=&0,~~~~~~~~~ \\
\Phi_2
   \left(\frac{A_t^2}{f^2}+\frac{zf'-k^2z^2-2f+2}{z^2f}+2\left(\frac{\Phi'}{\Phi}\right)^2+\frac{f' \Phi'}{f \Phi }-\frac{2 \Phi
   ^2}{f} \right)-\frac{4 A_t \Phi^2
  }{f} \Phi_1-\frac{2 \Phi ' }{\Phi }\Phi_2'+\Phi_2''&=&0.\label{w0scalar2}
\ee
in which $\Phi_3$ is also decoupled.

After solving the  EoMs \eqref{w0scalar1} and \eqref{w0scalar2} one can get a series QNMs of $k$. The correlation length $\xi=1/|{\rm Im}(k^*)|$ where $k^*$ is the lowest mode whose imaginary part is closest to the real axis. In Fig.\ref{thermal_xi}(a) we show the correlation length to the temperature both in normal state and superconducting state. We can see explicitly that $\xi\propto |1-T/T_c|^{-1/2}$ in both phases, which indicates that $\nu=1/2$ and consistent with mean field theory. The prefactors for the normal state is $\xi_0^>\approx0.5767/T_c\approx4.8708$ while the one in the superconducting state is $\xi_0^<\approx 0.6195/T_c\approx 5.2323$. Thus in the holographic model $\xi_0^><\xi_0^<$, which indicates a non-Ginzburg-Landau model.

The relaxation time $\tau$ can be obtained similarly.  By solving the EoMs \eqref{k0scalar1} and \eqref{k0scalar2}, one can get a series modes of $\omega$. Then $\tau=1/|{\rm Im}(\omega^*)|$ where $\omega^*$ is the lowest mode in those QNMs.  We can see explicitly from  Fig.\ref{thermal_xi}(b) that $\tau\propto |1-T/T_c|^{-1}$ for both normal and superconducting phases, which indicates that $z=2$ and consistent with the mean field theory. The prefactors in the normal state is $\tau_0^>\approx1.662/T_c\approx14.0372$ while the one in the superconducting state is $\tau_0^<\approx0.9574/T_c\approx8.0862$. The two prefactors $\tau_0^>$ and $\tau_0^<$ have large discrepancy compared to Ginzburg-Landau model, because we get them by integrating the fields in the whole bulk, different from direct calculations of them in the boundary like in Ginzburg-Landau theory. It also implies that although the boundary is like a mean field theory, they are not Ginzburg-Landau type.


\begin{thebibliography}{99}\footnotesize
\bibitem{sachdev}
S. Sachdev, ``Quantum phase transitions'', Second Edition, Cambridge University Press, (2011).

\bibitem{Maldacena:1997re}
  J.~M.~Maldacena,
  ``The Large N limit of superconformal field theories and supergravity,''
  Int.\ J.\ Theor.\ Phys.\  {\bf 38}, 1113 (1999)

\bibitem{Kibble:1976sj}
  T.~W.~B.~Kibble,
  ``Topology of Cosmic Domains and Strings,''
  J.\ Phys.\ A {\bf 9}, 1387 (1976);
  
 \bibitem{Kibble:1980mv}
  T.~W.~B.~Kibble,
  ``Some Implications of a Cosmological Phase Transition,''
  Phys.\ Rept.\  {\bf 67} (1980) 183.

\bibitem{Zurek:1985qw}
  W.~H.~Zurek,
  ``Cosmological Experiments in Superfluid Helium?,''
  Nature {\bf 317} (1985) 505;
\bibitem{Zurek:1996sj}
  W.~H.~Zurek,
  ``Cosmological experiments in condensed matter systems,''
  Phys.\ Rept.\  {\bf 276}, 177 (1996)
  [cond-mat/9607135].

\bibitem{Laguna:1996pv}
  P.~Laguna and W.~H.~Zurek,
  ``Density of kinks after a quench: When symmetry breaks, how big are the pieces?,''
  Phys.\ Rev.\ Lett.\  {\bf 78}, 2519 (1997)
  [gr-qc/9607041].

\bibitem{Yates:1998kx}
  A.~Yates and W.~H.~Zurek,
  ``Vortex formation in two-dimensions: When symmetry breaks, how big are the pieces?,''
  Phys.\ Rev.\ Lett.\  {\bf 80} (1998) 5477
  [hep-ph/9801223].

\bibitem{Ibaceta:1998yy}
  D.~Ibaceta and E.~Calzetta,
  ``Counting defects in an instantaneous quench,''
  Phys.\ Rev.\ E {\bf 60} (1999) 2999
  [hep-ph/9810301].

\bibitem{Antunes:1998rz}
  N.~D.~Antunes, L.~M.~A.~Bettencourt and W.~H.~Zurek,
  ``Vortex string formation in a 3-D U(1) temperature quench,''
  Phys.\ Rev.\ Lett.\  {\bf 82}, 2824 (1999)
  [hep-ph/9811426].




  \bibitem{donaire2007}
M. Donaire, T.W.B. Kibble and A. Rajantie, ``Spontaneous vortex formation on a superconducting film", New J. Phys. {\bf 9} (2007) 148.

\bibitem{Das:2011cx} 
  A.~Das, J.~Sabbatini and W.~H.~Zurek,
  ``Winding up superfluid in a torus via Bose Einstein condensation,''
  Sci.\ Rep.\  {\bf 2}, 352 (2011)
  [arXiv:1102.5474 [cond-mat.other]].

\bibitem{Gillman:2017ycq}
  E.~Gillman and A.~Rajantie,
  ``Kibble Zurek mechanism of topological defect formation in quantum field theory with matrix product states,''
  Phys.\ Rev.\ D {\bf 97} (2018) no.9,  094505
  [arXiv:1711.10452 [quant-ph]].








\bibitem{Chuang:1991zz}
  I.~Chuang, B.~Yurke, R.~Durrer and N.~Turok,
  ``Cosmology in the Laboratory: Defect Dynamics in Liquid Crystals,''
  Science {\bf 251} (1991) 1336.

\bibitem{Bowick:1992rz}
  M.~J.~Bowick, L.~Chandar, E.~A.~Schiff and A.~M.~Srivastava,
  ``The Cosmological Kibble mechanism in the laboratory: String formation in liquid crystals,''
  Science {\bf 263} (1994) 943
  [hep-ph/9208233].

\bibitem{Digal:1998ak}
  S.~Digal, R.~Ray and A.~M.~Srivastava,
  ``Observing correlated production of defect - anti-defects in liquid crystals,''
  Phys.\ Rev.\ Lett.\  {\bf 83} (1999) 5030
  [hep-ph/9805502].

\bibitem{Baeuerle:1996zz}
  C.~Baeuerle, Y.~M.~Bunkov, S.~N.~Fisher, H.~Godfrin and G.~R.~Pickett,
  ``Laboratory simulation of cosmic string formation in the early Universe using superfluid He-3,''
  Nature {\bf 382} (1996) 332.

\bibitem{Ruutu:1995qz}
  V.~M.~H.~Ruutu {\it et al.},
  ``Big bang simulation in superfluid He-3-b: Vortex nucleation in neutron irradiated superflow,''
  Nature {\bf 382} (1996) 334
  [cond-mat/9512117].






\bibitem{Carmi:2000zz}
  R.~Carmi, E.~Polturak and G.~Koren,
  ``Observation of Spontaneous Flux Generation in a Multi-Josephson-Junction Loop,''
  Phys.\ Rev.\ Lett.\  {\bf 84} (2000) 4966.

\bibitem{Monaco:2002zz}
  R.~Monaco, J.~Mygind and R.~J.~Rivers,
  ``Zurek-Kibble Domain Structures: The Dynamics of Spontaneous Vortex Formation in Annular Josephson Tunnel Junctions,''
  Phys.\ Rev.\ Lett.\  {\bf 89} (2002) 080603.

\bibitem{Monaco:2003}
  R.~Monaco, J.~Mygind, and R.~J.~Rivers,
   ``Spontaneous fluxon formation in annular Josephson tunnel junctions,''
   Phys.\ Rev.\ B\ {\bf 67} (2003) 104506.

\bibitem{Monaco:2005fi}
  R.~Monaco, J.~Mygind, M.~Aaroe, R.~J.~Rivers and V.~P.~Koshelets,
  ``Zurek-Kibble Mechanism for the Spontaneous Vortex Formation in Nb - Al/Al(ox)/Nb Josephson Tunnel Junctions: New Theory and Experiment,''
  Phys.\ Rev.\ Lett.\  {\bf 96} (2006) 180604
  [cond-mat/0503707 [cond-mat.supr-con]].

\bibitem{Maniv:2003zz}
  A.~Maniv, E.~Polturak and G.~Koren,
  ``Observation of Magnetic Flux Generated Spontaneously During a Rapid Quench of Superconducting Films,''
  Phys.\ Rev.\ Lett.\  {\bf 91} (2003) 197001.

  \bibitem{golubchik2010}
D. Golubchik, E. Polturak, G. Koren, ``Evidence for Long-Range Correlations within Arrays of Spontaneously Created Magnetic Vortices in a Nb Thin-Film Superconductor,'' Phys. Rev. Lett. {\bf 104}, 247002 (2010).

 \bibitem{golubchik2011}
D. Golubchik, E. Polturak, G. Koren, B. Ya. Shapiro and I. Shapiro, ``Experimental determination of correlations between spontaneously formed vortices in a superconductor,'' J Low Temp Phys (2011) 164: 74. [arXiv: 1101.0409]

\bibitem{Guo}
Xiao-Ye Xu {\it et al.}, ``Quantum Simulation of Landau-Zener Model Dynamics Supporting the Kibble-Zurek Mechanism,'' Phys. Rev. Lett. {\bf 112}, 035701(2014).

\bibitem{Kibble:2007zz}
  T.~Kibble,
  ``Phase-transition dynamics in the lab and the universe,''
  Phys.\ Today {\bf 60N9} (2007) 47.

\bibitem{delCampo:2013nla}
  A.~del Campo and W.~H.~Zurek,
  ``Universality of phase transition dynamics: Topological Defects from Symmetry Breaking,''
  Int.\ J.\ Mod.\ Phys.\ A {\bf 29} (2014) no.8,  1430018
  [arXiv:1310.1600 [cond-mat.stat-mech]].



\bibitem{Cubrovic:2009ye}
  M.~Cubrovic, J.~Zaanen and K.~Schalm,
  ``String Theory, Quantum Phase Transitions and the Emergent Fermi-Liquid,''
  Science {\bf 325}, 439 (2009)
  [arXiv:0904.1993 [hep-th]].

\bibitem{Adams:2012pj}
  A.~Adams, P.~M.~Chesler and H.~Liu,
  ``Holographic Vortex Liquids and Superfluid Turbulence,''
  Science {\bf 341}, 368 (2013)
  [arXiv:1212.0281 [hep-th]].

\bibitem{Witczak-Krempa:2013nua}
  W.~Witczak-Krempa, E.~Sorensen and S.~Sachdev,
  ``The dynamics of quantum criticality via Quantum Monte Carlo and holography,''
  Nature Phys.\  {\bf 10}, 361 (2014)
  [arXiv:1309.2941 [cond-mat.str-el]].

\bibitem{Bhaseen:2013ypa}
  M.~J.~Bhaseen, B.~Doyon, A.~Lucas and K.~Schalm,
  ``Far from equilibrium energy flow in quantum critical systems,''
  Nature Phys.\  {\bf 11} (2015) 5
  [arXiv:1311.3655 [hep-th]].


\bibitem{Zaanen:2015oix}
  J.~Zaanen, Y.~W.~Sun, Y.~Liu and K.~Schalm,
  ``Holographic Duality in Condensed Matter Physics,''
  Cambridge University Press, 2015

\bibitem{Chesler:2014gya}
  P.~M.~Chesler, A.~M.~Garcia-Garcia and H.~Liu,
  ``Defect Formation beyond Kibble-Zurek Mechanism and Holography,''
  Phys.\ Rev.\ X {\bf 5} (2015) no.2,  021015
  [arXiv:1407.1862 [hep-th]].

\bibitem{Sonner:2014tca}
  J.~Sonner, A.~del Campo and W.~H.~Zurek,
  ``Universal far-from-equilibrium Dynamics of a Holographic Superconductor,''
  Nature Commun.\  {\bf 6} (2015) 7406
  [arXiv:1406.2329 [hep-th]].

\bibitem{Das:2014lda}
  S.~R.~Das and T.~Morita,
  ``Kibble-Zurek Scaling in Holographic Quantum Quench : Backreaction,''
  JHEP {\bf 1501}, 084 (2015)
  [arXiv:1409.7361 [hep-th]].

\bibitem{Natsuume:2017jmu}
  M.~Natsuume and T.~Okamura,
  ``Kibble-Zurek scaling in holography,''
  Phys.\ Rev.\ D {\bf 95} (2017) no.10,  106009
  [arXiv:1703.00933 [hep-th]].

    \bibitem{tinkham}
  M. Tinkham, ``Introduction to Superconductivity", 2nd Edition, McGraw-Hill Inc. press (1996).
  
\bibitem{Francuz:2015zva}
  A.~Francuz, J.~Dziarmaga, B.~Gardas and W.~H.~Zurek,
  ``Space and time renormalization in phase transition dynamics,''
  Phys.\ Rev.\ B {\bf 93}, no. 7, 075134 (2016)
  [arXiv:1510.06132 [cond-mat.stat-mech]].
  
\bibitem{Sadhukhan:2019jan}
D.~Sadhukhan {\it et al.},
``Sonic horizons and causality in the phase transition dynamics,''
Phys. Rev. B \textbf{101} (2020) no.14, 144429
[arXiv:1912.02815 [quant-ph]].
  


  
\bibitem{Zurek:2013qba}
  W.~H.~Zurek,
  ``Topological relics of symmetry breaking: Winding numbers and scaling tilts from random vortex-antivortex pairs,''
  J.\ Phys.\ Condens.\ Matter {\bf 25}, no. 40, 404209 (2013)
  [arXiv:1305.4695 [cond-mat.stat-mech]].
  
  
\bibitem{Lin:2015lca}
  S.~Z.~Lin {\it et al.},
  ``Topological defects as relics of emergent continuous symmetry and Higgs condensation of disorder in ferroelectrics,''
  Nature Phys.\  {\bf 10}, 970 (2014)
  [arXiv:1506.05021 [cond-mat.mtrl-sci]].
  
    \bibitem{halperin}
 B. I. Halperin, Physics of Defects, Proceedings of the Les Houches Summer Institute (North Holland, Amsterdam, 1981).

 \bibitem{liu-mazenko}
F. Liu and G. F. Mazenko, ``Defect-defect correlation in the dynamics of first-order phase transitions'', Phys. Rev. B 46, 5963 (1992)
  

    \bibitem{hartnoll}
  S.~A.~Hartnoll, C.~P.~Herzog and G.~T.~Horowitz,
  ``Building a Holographic Superconductor,''
  Phys.\ Rev.\ Lett.\  {\bf 101} (2008) 031601
  [arXiv:0803.3295 [hep-th]].




\bibitem{Skenderis:2002wp}
  K.~Skenderis,
  ``Lecture notes on holographic renormalization,''
  Class.\ Quant.\ Grav.\  {\bf 19}, 5849 (2002)
  [hep-th/0209067].

  \bibitem{witten}
  E.~Witten,
  ``SL(2,Z) action on three-dimensional conformal field theories with Abelian symmetry,''
  In *Shifman, M. (ed.) et al.: From fields to strings, vol. 2* 1173-1200
  [hep-th/0307041].

\bibitem{silva}
  O.~Domenech, M.~Montull, A.~Pomarol, A.~Salvio and P.~J.~Silva,
  ``Emergent Gauge Fields in Holographic Superconductors,''
  JHEP {\bf 1008} (2010) 033
  [arXiv:1005.1776 [hep-th]].


\bibitem{Chesler:2013lia}
  P.~M.~Chesler and L.~G.~Yaffe,
  ``Numerical solution of gravitational dynamics in asymptotically anti-de Sitter spacetimes,''
  JHEP {\bf 1407} (2014) 086
  [arXiv:1309.1439 [hep-th]].




















































\end{thebibliography}

\begin{thebibliography}{99}\footnotesize

\bibitem{Kovtun:2005ev}
  P.~K.~Kovtun and A.~O.~Starinets,
  ``Quasinormal modes and holography,''
  Phys.\ Rev.\ D {\bf 72} (2005) 086009
  [hep-th/0506184].

\bibitem{Maeda:2009wv}
  K.~Maeda, M.~Natsuume and T.~Okamura,
  ``Universality class of holographic superconductors,''
  Phys.\ Rev.\ D {\bf 79} (2009) 126004
  [arXiv:0904.1914 [hep-th]].

\bibitem{Amado:2009ts}
  I.~Amado, M.~Kaminski and K.~Landsteiner,
  ``Hydrodynamics of Holographic Superconductors,''
  JHEP {\bf 0905}, 021 (2009)
  [arXiv:0903.2209 [hep-th]].
\end{thebibliography}
\end{document}